\documentclass[11pt]{article}
\usepackage{amsfonts,amsmath,amssymb,amsthm}
\usepackage{thmtools, thm-restate}
\usepackage{graphicx} 
\usepackage{float}
\usepackage{caption}
\usepackage{subcaption}
\usepackage{comment}

\usepackage{fullpage}
\usepackage[margin=0.5in]{geometry}
\usepackage{xcolor}
\usepackage{hyperref}
\usepackage[capitalize]{cleveref}
\usepackage{authblk}

\usepackage{titlesec}
\newcommand\largesubsection{%
  \titleformat{\section}
    {\normalfont\huge\bfseries\filcenter}{\thesection}{1em}{}
}
\newcommand\stdsubsection{%
  \titleformat{\section}
    {\normalfont\Large\bfseries}{\thesection}{1em}{}
}

\newtheorem{thm}{Theorem}
\newtheorem{theorem}{Theorem}
\newtheorem{defn}{Definition}

\newtheorem*{rem}{Remark}

\newenvironment{myproof}{
  \par\medskip\noindent
  \textit{Proof}.
}{
\newline
\rightline{$\qedsymbol$}
}

\def\eps{\varepsilon}
\def\E{\mathbb{E}}

\def\HP{\operatorname{HP}}
\def\R{\operatorname{R}}
\def\TO{\operatorname{TO}}
\def\L{\operatorname{L}}
\def\T{\operatorname{T}}
\def\TL{\operatorname{T_L}}

\def\B{\operatorname{B}}
\def\K{\operatorname{K}}
\def\S{\operatorname{S}}
\def\D{\operatorname{D}}
\def\C{\operatorname{C}}
\def\H{\operatorname{H}}

\def\deg{\operatorname{deg}}

\def\SI{Supplementary Information}

\def\maintitle{Colonization times in Moran process on graphs}
\title{\maintitle}

\author[1]{Lenka Kopfová} 
\author[1]{Josef Tkadlec}

\affil[1]{Computer Science Institute, Charles University, Prague, Czech Republic}

\date{}

\begin{document}


\maketitle

\begin{abstract}
Moran Birth-death process is a standard stochastic process that is used to model natural selection in spatially structured populations.
A newly occurring mutation that invades a population of residents can either fixate on the whole population or it can go extinct due to random drift.
The duration of the process depends not only on the total population size $n$,
but also on the spatial structure of the population.
In this work, we consider the Moran process with a single type of individuals who invade and colonize an otherwise empty environment.
Mathematically, this corresponds to the setting where the residents have zero reproduction rate, thus they never reproduce.
We present two main contributions.
First, in contrast to the Moran process in which residents do reproduce, we show that the colonization time is always at most a polynomial function of the population size $n$. Namely, we show that colonization always takes at most $\frac12n^3-\frac12n^2$ expected steps, and for each $n$, we exactly identify the unique slowest spatial structure where it takes exactly that many steps.
Moreover, we establish a stronger bound of roughly $n^{2.5}$ steps for spatial structures that contain only two-way connections and an even stronger bound of roughly $n^2$ steps for lattice-like spatial structures.
Second, we discuss various complications that one faces when attempting to measure fixation times and colonization times in spatially structured populations, and we propose to measure the real duration of the process, rather than counting the steps of the classic Moran process.
\end{abstract}

\section*{Introduction}
Natural selection is a stochastic process that acts on populations of reproducing individuals~\cite{kimura1968evolutionary,burger2000mathematical,ewens2004mathematical}.
As time goes by, individuals acquire mutations that affect their reproductive rate.
The advantageous mutations generally tend to propagate through the population,
whereas the frequency of disadvantageous mutations tends to go down.
When mutations are sufficiently rare, the key question is to determine the fate of a single newly occurring mutation as it attempts to invade a homogeneous population of residents.
This fate depends on several factors, such as the population size $n$ or the relative fitness advantage $r$ that the mutation grants onto its bearer.

Another important factor that greatly affects the evolutionary dynamics is the spatial structure of the population~\cite{durrett1994importance,frean2013effect}.
Evolutionary graph theory is a framework developed to study those effects~\cite{lieberman2005evolutionary}.
The individuals are represented as nodes of a graph (network), and the connections between the nodes represent the possible migration patterns. The connections can be one-way or two-way.
Graphs can represent arbitrary spatial structures, including well-mixed populations, metapopulations~\cite{yagoobi2021fixation,yagoobi2023categorizing,svoboda2023coexistence,svoboda2024amplifiers}, or lattices~\cite{kaveh2015duality,tkadlec2023evolutionary}.
The evolutionary dynamics is governed by the Moran Birth-death process~\cite{moran1958random,lieberman2005evolutionary}.
That is, in each step, first an individual is selected for reproduction with probability proportional to its fitness, and then the offspring migrates and replaces a random neighbor, see~\cref{fig:moran}.

\begin{figure}[h]
  \centering
   \includegraphics[width=1\linewidth]{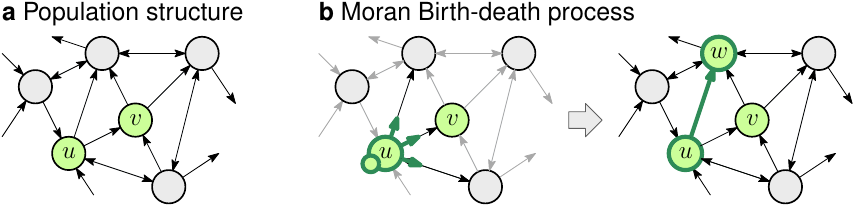}
\caption{
\textbf{Moran Birth-death process on a spatial structure.}
\textbf{a,} The spatial structure is given as a network (graph), where nodes represent sites and arrows represent possible migration patterns. Nodes occupied by mutants are green (here $u$ and~$v$).
\textbf{b,} In each step of the Moran process, first, a random node is selected for reproduction, and then the offspring migrates along a random outgoing edge. Here, the offspring of $u$ migrated to $w$.
}
\label{fig:moran}
\end{figure}

Two key quantities that describe the fate of a new mutation are fixation probability and fixation time~\cite{kimura1962probability,kimura1980average,slatkin1981fixation,whitlock2003fixation}.
Fixation probability is the chance that the mutation eventually spreads throughout the population.
Fixation time is the number of steps of the Moran process until this happens, and it captures the duration of the process.
Both quantities have been studied extensively~\cite{broom2008analysis,hadjichrysanthou2011evolutionary,monk2014martingales,diaz2014approximating,pavlogiannis2017amplification,moller2019exploring,tkadlec2019population,monk2020wald}.
For example, it is known that certain spatial structures dramatically increase the fixation probability of even mildly advantageous mutations~\cite{galanis2017amplifiers,pavlogiannis2018construction,goldberg2019asymptotically}, though such a boost must always come at the cost of an increase in fixation time~\cite{tkadlec2021fast}.

The fixation time crucially depends on the spatial structure of the population.
For example, when the initial mutant has a constant fitness advantage $r>1$, the number of steps is
roughly $n\log n$ for the well-mixed population~\cite{nowak2006evolutionary},
roughly $n^2$ for a population arranged along a cycle~\cite{hathcock2019fitness}, and
roughly $n^3$ for a population organized as a so-called double star~\cite{goldberg2020phase}.
The double stars are known to be essentially the slowest possible structures among those where all connections are two-way~\cite{goldberg2020phase}.
However, when some connections are one-way, the fixation time can be as large as exponential in $n$~\cite{diaz2016absorption}.
In general, no efficient algorithm is known to compute or even approximate the fixation time on a given spatial structure that contains one-way edges.

Given those difficulties, Moran process on population structures is often studied in different limits that make the analysis more tractable.
The limit $r\to 1$ is called \textit{weak selection}.
It corresponds to settings where the mutation grants only a marginal advantage.
That is, the invading mutants reproduce only barely more frequently than the existing residents.
Evolutionary dynamics under weak selection can be approximated for any population structure~\cite{allen2017evolutionary,mcavoy2021fixation,durocher2022invasion,brendborg2022fixation}.
Moreover, formulas for fixation times are known~\cite{altrock2009fixation,gao2024speed}.

In the opposite limit, mutants reproduce at a much higher rate than the residents.
This limit is called the \textit{ecological scenario}~\cite{ibsen2015computational}.
It has been studied, for instance,
in the context of death-Birth updating~\cite{allen2020transient,tkadlec2020limits},
in order to obtain approximations for the fixation probabilities~\cite{brewster2024fixation},
or when analyzing range expansion~\cite{komarova2022laws}.
Mathematically, the ecological scenario corresponds to the setting 
in which residents have a reproductive rate 0, thus the mutant relative fitness advantage $r$ satisfies  $r\to\infty$.
Biologically, this regime thus models situations such as a new invasive species colonizing an initially empty spatially structured environment.

In the ecological scenario, mutants eventually expand to all reachable parts of the environment.
The \textit{colonization time} is the expected number of steps until this happens.
That is, colonization time is a direct analogue of the fixation time in the limit $r\to\infty$.

In this work, we consider colonization times on arbitrary spatial structures, with or without one-way edges.
As our main theoretical result, for every population size $n$ we precisely pinpoint the unique population structure with the slowest colonization time,
and we show that this slowest colonization time is of the order of $n^3$ steps.
Thus, while fixation times on some spatial structures with fixed $r>1$ may be exponentially long, colonization time on any spatial structure is always at most polynomial.
Moreover, we present a stronger bound of $n^{2.5}$ steps for those spatial structures in which all connections are two-way, and an even stronger bound $n^2$ steps for those spatial structures in which each node in the network has the same number of neighbors.
To conclude, we discuss and compare several possible ways to measure colonization times and fixation times in spatially structured populations.


\section*{Model}
In this section, we describe the notions of our model in detail.

\subsection*{Spatial structure}
The population structure is represented by a directed graph $G$.
The nodes of $G$ represent individuals, and the graph edges correspond to the connections between them.
At any point in time, each node is either a \textit{mutant} or a \textit{resident}.
Initially, there is only a single mutant at node $v$.
We require that there exists a directed path from $v$ to any other node.
This guarantees that in the limit $r\to\infty$ the process terminates with mutant fixation, with probability 1 in finite expected time.
At any given time, we denote by $M$ the set of nodes that are mutants, and we refer to the tuple $(G,M)$ denoting the graph and its current set of mutants as a \emph{state} of the process.
A special case of graphs we concentrate on is the class of \textit{undirected} graphs in which all connections are two-way.
For a vertex $v$, we define its \emph{(out)degree} denoted as $\deg(v)$ to be the number of outgoing edges incident with $v$.
An even more special case is the class of \textit{regular} graphs in which all vertices have the same degree. This class includes, for example, lattices of any connectivity.

\subsection*{Moran process at $r\to\infty$}
We consider a modified version of the classic Moran birth-death process in which only mutants reproduce. In other words, mutants have fitness 1 and residents have fitness 0, thus the relative mutant advantage is $r\to\infty$. This process models situations in which an invading type colonizes an otherwise empty environment. In each step of this modified Moran process, we first pick a uniformly
random node. If the node is a mutant, it reproduces onto a random neighbor; otherwise, nothing happens. More formally, suppose we pick a node $u$:
\begin{enumerate}
    \item  If $u$ is a mutant, we select another node $u'$ uniformly at random from among the $\deg(u)$ nodes connected to $u$, and we set $u'$ to be a mutant too. (Note that this changes the state of the population if and only if $u'$ used to be a resident.)
    \item If $u$ is a resident, no change occurs.
\end{enumerate}

\subsection*{Colonization time}
Given a graph structure $G$ and a starting node $v$, we define the colonization time $\T(G,v)$ to be the expected number of steps until mutants fixate when the modified Moran Birth-death process is run on the graph $G$.
We will often study the worst-case scenarios in order to prove upper bounds on the quantity $\T(G,v)$. Thus, we also denote by $\T(G) = \max_{v\in V(G)}\T(G,v)$ the maximum expected colonization time over all possible starting nodes $v$.
Note that the colonization time accounts for all the steps of the process, including those steps in which a resident node was picked and did not reproduce. While perhaps counter-intuitive at first, this way of measuring time turns out to better correspond to the ``real'' duration of the process. See section Discussion, for an in-depth discussion of the connections among different ways of measuring time.




\subsection*{Asymptotic notation}
Throughout this text, we use the asymptotic notation $o(\cdot)$, $O(\cdot)$, $\Omega(\cdot)$ and $\Theta(\cdot)$ to denote that some function $f$ is asymptotically strictly smaller than some other function $g$ (denoted $f=o(g)$), asymptotically smaller than or equal to $g$ ($f=O(g)$), asymptotically larger than $g$ ($f=\Omega(g)$) and asymptotically equal to $g$ ($f=\Theta(g)$). We will also use the symbol $\approx$ to denote "approximately equal to," meaning $f(n)\approx g(n)$ if $f(n)=g(n)+o(g(n))$. For example $\frac{1}{2}n^2+3n = o(n^3)=O(n^2)=\Omega(n\log n)=\Theta(n^2).$
See~\cite{cormen2022introduction} and \SI{} for details.

\section*{Results}
Our main analytical results are bounds on the colonization time for different classes of graphs.
In particular, we show a general upper bound of $O(n^3)$ that applies to all spatial structures and we prove that it is exactly tight. Then, we proceed by improving this upper bound to $O(n^{2.5})$ for undirected graphs and to $O(n^2)$ for regular graphs. We also compute asymptotically precise colonization times for specific graph classes such as the complete graphs $K_n$, the cycle graphs $C_n$, the star graphs $S_n$, and other graphs.

\subsection*{General bounds}
First, we consider the general setting with no constraints on the population structure. We prove that, in this case, the colonization time is always at most cubic in the population size $n$.
\begin{theorem}[General upper  bound]\label{thm:directed_upper_bound} 
Let $G_n$ be a graph (directed or undirected) with $n$ nodes.
Then $\T(G_n)\leq\frac12 n^3-\frac{1}{2}n^2$.
\end{theorem}

In particular, the colonization time on any population structure is at most polynomial in the population size $n$. Note that this contrasts with the regime of finite $r>1$. In that regime, the fixation time on some spatial structures is known to be exponential~\cite{diaz2016absorption}.

The idea behind the proof is to decompose the process into stages such that each stage lasts until we gain a new mutant.
We then argue that for any spatial structure, each individual stage can take at most $O(n^2)$ steps on average. Since in total, there are $n-1$ stages, by linearity of expectation this gives a cubic upper bound for the total number of steps.
See \SI{} for details.

\begin{figure}[h]
  \centering
  \includegraphics[width=\linewidth]{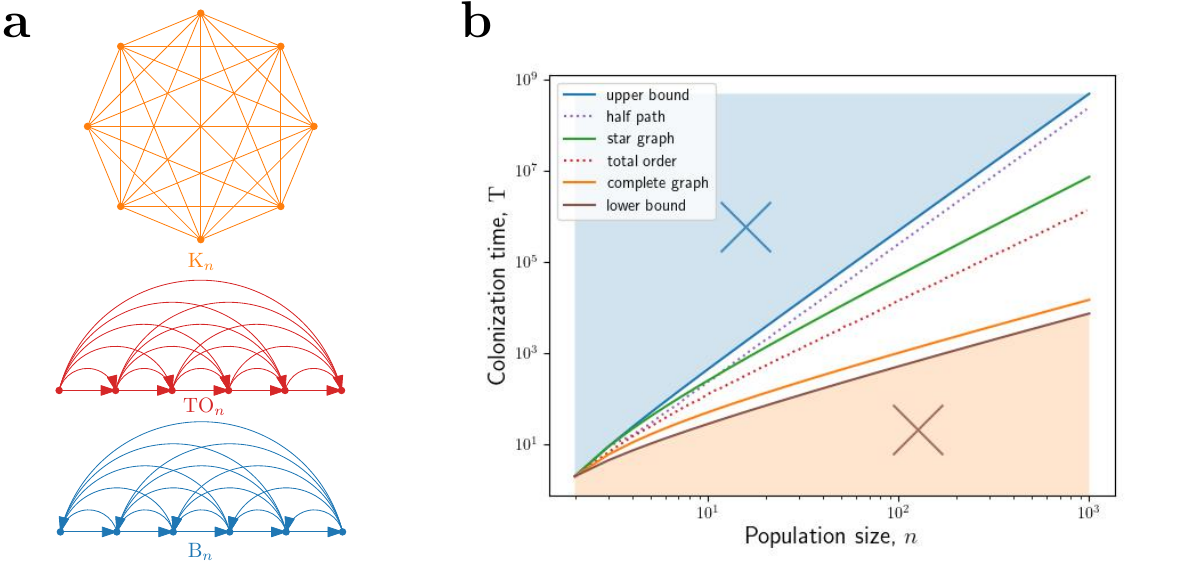}
\caption{
\textbf{Colonization times on directed graphs.}
\textbf{a,} In the complete graph $\K_n$, each two nodes are connected by a two-way edge. In the total order graph $\TO_n$, the nodes are arranged left to right and all edges going left to right are included.  The backward graph $\B_n$ consists of a directed path going left to right, plus all one-way edges going in the opposite direction, right to left. 
\textbf{b,} For each $n$, the backward graph $\B_n$ (blue) is the graph with maximal colonization time. We have $\T(\B_n)=\frac12n^3-\frac12n^2$. The shortest possible colonization time is of the order of $n\log n$ steps, which is achieved for the complete graph $\K_n$ (orange). For the total order graph $\TO_n$ (red) we have $\T(\TO_n)= \Theta(n^2)$. Here the lines show the proved analytical results, the dots show the simulations, and the axes are log-scale.
}
\label{fig:directed}
\end{figure}

Somewhat surprisingly, we show that the upper bound in~\cref{thm:directed_upper_bound} is exactly tight. That is, we identify a population structure which we call a \textit{backward graph} $\B_n$ for which the bound is achieved with equality, see~\cref{fig:directed}a.

\begin{theorem} 
For every $n$ there exists a directed graph $\B_n$ and an initial mutant node $v$ of $\B_n$ such that $\T(\B_n,v) = \frac12 n^3-\frac{1}{2}n^2$.
\end{theorem}

To sum up, the longest possible colonization time on any population structure is equal to $\frac12n^3-\frac12n^2$. We also show that the shortest possible colonization time is of the order of at least $n\log n$ steps and that this is the case e.g.\ for the complete graph $\K_n$. See Theorems~5 and~11 in the 
\SI{} for details. Finally, we show that the colonization time on a so-called total order graph $\TO_n$ is of the order of $n^2$ steps, see~\cref{fig:directed}b. 




\subsection*{Stronger bound for undirected graphs}
Our first result shows that the colonization times range from roughly $n\log n$ steps to roughly $n^3$ steps. Here we show that when the graph is undirected, that is, all connections are two-way, then the upper bound can be improved to roughly $n^{2.5}$ steps.

\begin{theorem}\label{thm:undirected} Let $G_n$ be an undirected graph with $n$ nodes. Then $\T(G_n) \leq 4n^2\sqrt{n}+o(n^2\sqrt{n})= O(n^2\sqrt{n})$.
\end{theorem}

\begin{figure}[h]
  \centering
   \includegraphics[width=\linewidth]{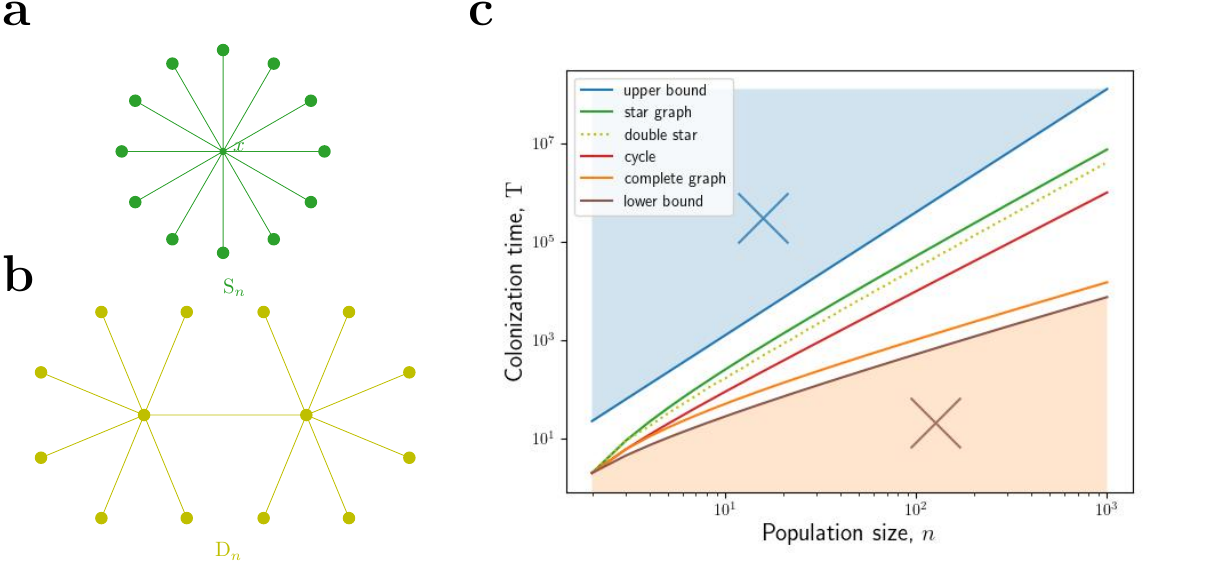}
\caption{\textbf{Colonization times on undirected graphs.}
\textbf{a,} In the star graph $\S_n$, one node is the center, and all the other nodes are connected to it by a two-way edge. The double star graph $\D_{2k}$ is obtained by joining the centers of two star graphs $\S_{k}$ using a two-way edge.
\textbf{b,} The proved upper bound for undirected graphs is $4n^2\sqrt{n}+o(n^2\sqrt{n})$. Here we plot the function $4n^2\sqrt{n}$ with blue color. The graph with the largest colonization time we found is the star graph $\S_n$ (green). Again the shortest possible colonization time is of the order of $n\log n$ steps, which is achieved for the complete graph $\K_n$ (orange). For the double star graph $\D_n$ (yellow) we have $\T(\D_n)= \Theta(n^2\log{n})$. For the cycle $\C_n$ (red) we have $\T(\C_n)=\Theta(n^2)$. Here the lines show the proved analytical results, the dots show the  simulations, and the axes are log-scale.
}
\label{fig:undirected}
\end{figure}

Recall that in the regime of finite $r>1$, the undirected graph with the largest known fixation time is the so-called double star $\D_n$~\cite{goldberg2020phase}, see~\cref{fig:undirected}a. The fixation time is of the order of roughly $n^3$ steps. Thus, \cref{thm:undirected} shows that even in the special case of undirected graphs, the colonization times are generally substantially shorter than the fixation times.

The idea behind the proof is again to divide the process into stages. While a single stage can be relatively long, we are able to argue that any such ``long'' stage must be balanced off by several subsequent ``short'' stages. By amortization, we then prove that the stages take at most $O(n\sqrt{n})$ steps, on average. See \SI{} for details.



We also compute colonization times on several specific undirected spatial structures.
We show that for both the star $\S_n$ and the double star $\D_n$ the colonization time is of the order of $n^2\log n$ steps. In fact, we show that the colonization time on the star is a constant factor 
 larger than the time on the double star. This is in contrast to the regime of fixed $r>1$, where the fixation time on the star is also of the order of $n^2\log n$ steps~\cite{tkadlec2019population}, whereas the fixation time on the double star is of the order of roughly $n^3$ steps~\cite{goldberg2020phase}.
For the cycle graph $\C_n$, we show that the colonization time is of the order of $n^2$ steps, the same as its fixation time in the regime $r>1$. See~\cref{fig:directed} for an illustration and the \SI{} for details.




\subsection*{Even stronger bound for regular graphs}
Previous research has highlighted the importance of regular graphs, that is, graphs in which each node is connected to the same number of neighbors.
Such graphs are also sometimes called isothermal graphs, since in the neutral evolution, each node is on average replaced equally often by its neighbors.
The class of regular graphs includes lattice-like structures of any connectivity $d$.
The Isothermal theorem states that the fixation probability of a single mutant with relative fitness advantage $r>1$ is the same for all regular graphs~\cite{nowak2006evolutionary}.

As our final analytical result, for regular graphs we improve the upper bound on the colonization time to $O(n^2)$.

\begin{theorem}\label{thm:regular} Let $G_n$ be a regular undirected graph with $n$ nodes. Then $\T(G_n) = O(n^2).$ 
\end{theorem}

Note that the upper bound is asymptotically shorter than the colonization time on the star graph, which is of the order of $n^2\log n$. Thus, regular graphs generally have shorter colonization times than undirected graphs, which in turn have shorter colonization times than arbitrary graphs.

The idea behind the proof is similar to the proof of~\cref{thm:undirected}.
We again split the process into stages. Since the graph is regular, we are able to argue that any single ``long'' stage must be followed by ``many'' short stages. The amortization argument then gives a stronger upper bound as compared to the case of undirected graphs. See~\SI{} for details.


We note that the dependence on $n$ in~\cref{thm:regular} can not be improved. This is because for the cycle graph $C_n$ the colonization time is of the order of $n^2$ steps.

\section*{Discussion}
The fixation time of a newly occurring mutation is a key factor in evolutionary dynamics.
Apart from depending on the population size $n$, the fixation time also depends on the relative mutant fitness advantage $r$ and on the spatial structure of the population.
When the mutant fitness advantage $r>1$ is fixed, there exist large spatial structures for which the fixation time is exponentially large in the population size $n$~\cite{diaz2016absorption}.
However, as we show in this work, this kind of long-term coexistence of invading mutants and existing residents can not occur in the limit of large mutant fitness advantage $r\to\infty$, which corresponds to a species colonizing an empty environment.
In this regime, the colonization time on the slowest possible spatial structure, which we call a backward graph, is only $\frac12n^3-\frac12n^2$ steps. 

Existing literature in the field of evolutionary graph theory highlighted the role of spatial structures in which all connections are two-way~\cite{broom2008analysis,pavlogiannis2017amplification,goldberg2019asymptotically}.
Those structures are described by undirected graphs.
The slowest known undirected graphs are the so-called double stars $\D_n$.
For any fixed $r>1$, the expected fixation time on a double star $\D_n$ is roughly $n^3$ steps~\cite{goldberg2020phase}.
In contrast, here we show that in the regime $r\to \infty$, the expected fixation time on any undirected graph is of the order of at most $n^{2.5}$ steps.
In particular, the fixation time on double stars drops to roughly $n^2\log n$.
Moreover, we show that double stars cease to be the slowest graphs since (plain) stars are a constant factor slower (see \SI{} for details).
While stars are the slowest undirected graphs that we found, in principle there could exist undirected graphs with colonization times as long as $n^{2.5}$. Identifying the exact slowest undirected graphs is an interesting problem left for future work.
We note that any such graphs, if they exist at all, would have to be irregular, since for regular graphs we proved an even stronger upper bound of at most $n^2$ steps.

Rigorous analysis of fixation times on spatial structures is made difficult by several factors.
In what follows we elaborate on four of them.

First, the fixation time is not a number but a random variable (unlike e.g.\ the fixation probability).
That is, depending on what individuals are selected for reproduction at each step, the total number of steps could be very small or very large.
The standard approach to treat this is to study the  \textit{expected} fixation time, that is, to replace the random variable with its expectation (a number).
This is often quite sufficient since the random variable is typically well concentrated~\cite{diaz2014approximating,hathcock2019fitness}.
(We note that for special spatial structures such as cycles the full distribution of the fixation time is understood~\cite{hathcock2019fitness}.)

Second, there are in fact two competing notions of fixation time that differ in what evolutionary trajectories are taken into account.
One notion is the unconditional fixation time (also known as the absorption time) which averages over all evolutionary trajectories, regardless of whether the mutants have fixated or gone extinct.
Alternatively, one can consider the conditional fixation time which averages over only those trajectories in which the mutants have fixated.
The two times could be quite different.
For example, for a single mutant with fitness $r=1+\eps$ who is invading a large well-mixed population the absorption time is roughly $2n\log n$ steps, whereas the conditional fixation time is roughly $\frac2{\eps}\cdot n\log n$ steps~\cite[Theorem 4]{tkadlec2019population}.
When $\eps=0.01$, the second quantity is $100\times$ larger than the first one.
In this work, we deal with the limit $r\to\infty$ in which the two notions coincide, since all evolutionary trajectories terminate with the mutants fixating.

Third, the fixation time generally depends on the starting location of the mutant.
For example, on a large star graph $\S_n$ with $r>1$ fixed, the absorption time of a mutant starting at the center node is roughly $n\log n$ steps, whereas for the mutant starting at any of the leaves it is roughly $n^2\log n$ steps.
One natural approach to handle this is to average over the possible starting positions (either uniformly, or according to some distribution such as the so-called temperature~\cite{adlam2015amplifiers}).
Alternatively, as we do in this work, one can specify the starting node.

Fourth, one should specify the units in which the time is measured.
This issue is more subtle than it might seem at first glance.
In this work, we count steps of a certain slightly modified Moran process (see section Model).
However, by far the most popular approach is to count the steps of the classic Moran process~\cite{lieberman2005evolutionary} and, possibly, in the end normalize by a factor of $n$ to get to ``generations''~\cite{tkadlec2021fast}) to capture the fact that the $n$ individuals are reproducing in parallel.
The disadvantage of using the steps of the classic Moran process as a basis for measuring time is that it leads to certain counter-intuitive results, see e.g.~\cite{hindersin2014counterintuitive,diaz2016absorption}.

To present yet another paradoxical consequence of counting the steps of the classic Moran process, consider the population structured as a so-called lollipop graph $\L_n$ with $\sqrt n$ nodes along a directed path and the remaining nodes in a fully connected cluster (see~\cref{fig:lollipop}).
Biologically, such a structure could represent a stream leading to a pond.
If the initial mutant appears at the start of the path, the mutants eventually fixate with probability one -- they simply make their way along the path and then they repeatedly invade the cluster until one such attempt succeeds. On average, this happens after some number of steps.
However, if we initially place additional mutants in the cluster, then the expected number of steps may increase.
Intuitively, this is because mutants in the cluster are selected for reproduction more often than the residents would be, and this slows down the progress of the mutants along the path.
This effect becomes especially pronounced in the limit $r\to\infty$.
In \SI{} we show that with the first initial condition, the process terminates after roughly $n\log n$ expected Moran steps, whereas with the second one it terminates after roughly $n^{1.5}$ expected Moran steps. Since for large $n$, we have $n^{1.5}\gg n\log n$, we conclude that adding initial mutants might substantially increase the number of steps of the classic Moran process.

\begin{figure}[h]
  \centering
   \includegraphics[width=1\linewidth]{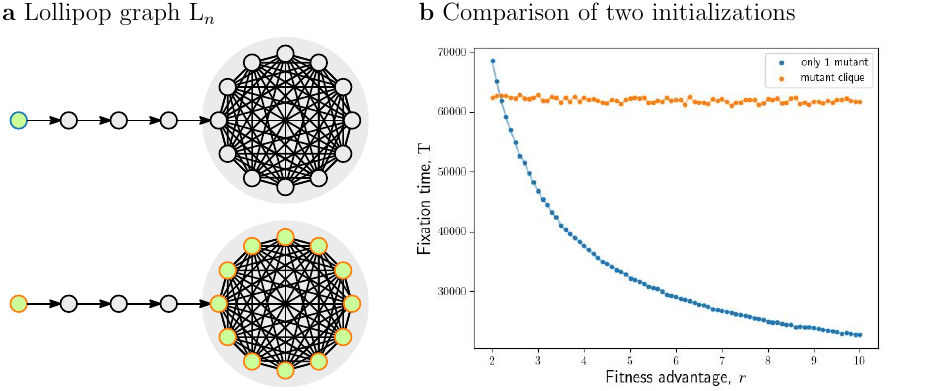}
\caption{
\textbf{Starting with more mutants causes more classic Moran steps.}
\textbf{a,} The lollipop graph $\L_n$ consists of $\sqrt n$ nodes arranged along a directed path, and the remaining nodes in a fully connected cluster (here $n=16$).
\textbf{b,} The classic fixation time on a lollipop graph $\L_n$ with two different initializations: either a single mutant at the start of the path (blue), or additionally all mutants in the fully connected cluster (orange). For $r>2.2$ 
 the first initialization leads to fewer classic Moran steps, despite having a strict subset of nodes that are initially mutants. Here $n=1600$, $r\in[2,10]$, and at least $10^3$ simulations per data point.
}
\label{fig:lollipop}
\end{figure}

To circumvent those paradoxical results, we propose to define the time units in a variable way depending on the total fitness of the population.
Formally, if the total fitness of the population is $F$, then we propose that one step of the classic Moran process accounts for $1/F$ units of ``real'' time.
Mathematically, this way of measuring time exactly corresponds to the situation in which Moran process is run in continuous time, and the reproduction time of any individual with fitness $f$ is an exponentially distributed random variable with parameter $1/f$~\cite{gillespie1977exact}.

In the case of constant selection ($r>1$ fixed), we have $n\le F\le r\cdot n$, therefore each step lasts at least $1/(rn)$ and at most $1/n$ units of real time.
Thus, up to a constant factor at most $r$, the real time corresponds to the the standard fixation time measured in generations (rather than in steps).

However, the difference might become much more pronounced in other regimes.
For example, consider again the lollipop graph $\L_n$ with initially a single mutant at the start of the path (that is, consider the first initial condition from~\cref{fig:lollipop}). Suppose that mutants have fitness 1 and that residents have fitness 0 (thus we are in the regime $r\to\infty$). In the~\SI{} we show that the classic Moran process then takes roughly $n\log n$ steps, which is roughly $\log n$ generations, whereas the real time is roughly $\sqrt n$ units. (Intuitively, this is because each of the $\sqrt n$ nodes along the directed path must become a mutant, and each one of them becomes a mutant after 1 unit of real time, on average.) Thus, neither classic Moran steps nor generations correctly represent the total duration of the process. On the other hand, we show that the colonization time, which is based on the modified Moran process (see section Model) is roughly $n\sqrt n$ steps, which is exactly a factor $n$ more than the real time. In fact, we show that this connection between the colonization time and the real time exists for any spatial structure: in order to compute the real time in the regime where mutants have fitness 1 and residents have fitness 0, one should compute the colonization time (in steps, based on the modified Moran process), and then divide by $n$.
This connection is the reason that in the limit $r\to\infty$ we work with the modified Moran process in the first place.
See~\SI{} for details.
To summarize, our bounds on colonization time yield the following bounds on real time: The colonization process terminates after at most $n^2$ units of real time on any spatial structure, after at most $n^{1.5}$ units of real time on any undirected structure, and after at most $n$ units of real time on any lattice-like structure.

\newpage
\largesubsection
\section*{Supplementary information}
\stdsubsection

This is a supplementary information to the manuscript \textit{\maintitle}.
It contains formal proofs of the theorems listed in the main text, namely:

\begin{restatable}{thm}{directedupperbound}\label{thm:upper bound}
Let $G_n$ be a graph (directed or undirected) with $n$ nodes.
Then $\T(G_n)\leq\frac12 n^3-\frac{1}{2}n^2$.
\end{restatable}

\begin{restatable}{thm}{backwardgraph}\label{thm:backward}
For every $n$ there exists a directed graph $\B_n$ and an initial mutant node $v$ of $\B_n$ such that $\T(\B_n,v) = \frac12 n^3-\frac{1}{2}n^2$.
\end{restatable}

\begin{restatable}{thm}{undirected}\label{thm:undirected_upper_bound}
Let $G_n$ be an undirected graph with $n$ nodes. Then $\T(G_n) \leq 4n^2\sqrt{n}+o(n^2\sqrt{n})= O(n^2\sqrt{n})$.
\end{restatable}

\begin{restatable}{thm}{regular}\label{thm:regular}
Let $G_n$ be a regular undirected graph with $n$ nodes. Then $\T(G_n) = O(n^2).$
\end{restatable}

\noindent The text is organized as follows:

In~\cref{sec:preliminaries} we introduce the formal notions, in \cref{sec:general} we prove the general upper $O(n^3)$ and lower bound $\Omega(n\log n)$ and show that the upper bound is tight (\cref{thm:upper bound} and \cref{thm:backward} from the main text), in \cref{sec:regular} we improve the upper bound to $O(n^2)$ for regular graphs (\cref{thm:regular} from the main text), in \cref{sec:undirected} we extend the previous ideas to get the upper bound $O(n^2\sqrt{n})$ for undirected graphs (\cref{thm:undirected_upper_bound} from the main text). In \cref{sec:specific} we analyze the specific graph families such as cycles, cliques, stars, double stars and total order graphs, and in \cref{sec:comparison} we illustrate various counter-intuitive properties of measuring time on the example of the lollipop graph.

\section{Preliminaries}\label{sec:preliminaries}
Our model consists of a graph $G$ (directed or undirected) representing a population structure. The vertices of $G$ denoted by $V(G)$ correspond to the individuals in the population. Two vertices directly interact if one is a neighbor of the other, meaning there is an edge between them. At any point, the individual associated with a given vertex can be either a mutant or a resident.
Let us call a setting where a particular subset of vertices are mutants as a \emph{state}.
More formally:

\begin{defn}
We call a tuple $(G, M)$ a \emph{state} where $G$ is a directed or undirected graph representing a particular population structure. And $M\subseteq V(G)$ is the set of mutant vertices. Then, $V(G)\setminus M$ is the set of resident vertices. We will usually denote $|V(G)|=n$ the number of individuals in our population.
\end{defn}

The population evolves in time, transiting to different states according to the so-called Moran birth-death process. Considering some fixed value $r$, every vertex has assigned a fitness value (the rate at which it reproduces) and mutants have relative fitness advantage $r.$ We can consider only cases when fitness is $r>1$ for mutants and $1$ for residents (regardless of the vertex position in the graph~$G$) up to scaling. We will refer to this Moran process with finite $r>1$ as a finite case. One step of the classical Moran birth-death process consists of two phases -- selection and reproduction (sometimes also called birth and death). In the selection phase, an individual $u$ for reproduction is selected proportionally to its fitness. Then, in the reproduction phase, it chooses a uniformly random neighbor $v$ in the graph $G$ into which it spreads. That means if $u$ is a mutant, then $v$ becomes a mutant, and if $u$ is a resident, then $v$ becomes a resident. Note that if both $u$ and $v$ are of the same type, the state doesn't change. Otherwise, the new state is the same except for exactly one vertex.

\begin{figure}[!hbt]
    \centering
    \includegraphics[scale=0.6]{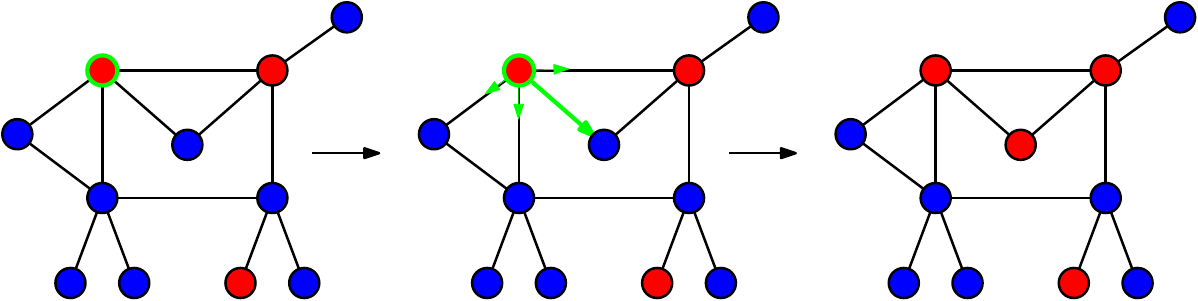}
    \caption{One step of the Moran process.}
    \label{fig:moran}
\end{figure}

These steps are repeated until the process reaches absorption, that is until all of the vertices become mutants (a state known as \emph{fixation}) or all of them become residents (a state known as \emph{extinction}). Note that on a general graph, the process may never be absorbed. However, we will consider only those population structures such that the absorption happens with probability $1$. Once the absorption is ensured, we may investigate some values associated with it. The probability that the process ends in fixation is called \emph{fixation probability}. We can also consider a random variable counting the number of steps until the process reaches absorption. The expected value of this variable is then called \emph{absorption time}. Similarly, we define the \emph{fixation time} as the expected value of the same random variable conditioned on the fixation event. That means we look only at those trajectories that end in fixation. 

\subsection*{Moran process at $r\to\infty$}

In this paper, we consider a modified version of the classical Moran birth-death process, which captures the idea of a strong selection of mutants, meaning that only mutants can reproduce. We consider the mutant fitness to be~$1$ and the resident fitness to be 0, thus the mutant advantage $r\to\infty.$

Looking at the Moran process at $r\to\infty$ corresponds to picking a random mutant uniformly at random which reproduces. More formally we can describe this process as:

\begin{defn} [Classic Moran process] 
Consider a given graph $G$ and its state $(G,M)$. One step of the \emph{classic Moran process} is defined as follows:
\begin{enumerate}
    \item (birth) A mutant vertex $m\in M$ is selected for reproduction with probability~$\frac{1}{|M|}$.
    \item (death) A neighbor $u$ of $m$ is selected uniformly at random and $m$ spreads into $u$. The probability of picking $u$ is thus $\frac{1}{\deg(m)}.$
\end{enumerate}

\end{defn}

Alternatively, we can consider a modified Moran process where we pick a uniformly random vertex, and then if it is a mutant, it reproduces; otherwise, we pick again.

\begin{defn} [Continuous process -- modified Moran process]
Consider a given graph $G$ and its state~$(G,M)$. One step of the \emph{continuous process} is defined as follows:
\begin{enumerate}
    \item (birth) A uniformly random vertex is selected. If it is a mutant, it reproduces and continues to the death phase. However, if it is a resident, nothing happens, and we will repeat this phase. Thus a particular mutant vertex $m\in M$ is selected for reproduction with probability $\frac{1}{|V(G)|}$ and the probability that some mutant vertex is selected is $\frac{|M|}{|V(G|}.$
    \item (death) A neighbor $u$ of $m$ is selected uniformly at random and $m$ spreads into $u$. The probability of picking a given neighbor $u$ of $m$ is thus $\frac{1}{\deg(m)}.$
\end{enumerate}

\end{defn}

In \cref{sec:comparison}, we argue why we think this modified second process is more natural, and thus, we will consider it to be the default one.

Note that in these processes, once a vertex becomes mutant, it stays mutant forever. Therefore, absorption occurs if and only if fixation occurs.
Hence, if some resident vertex is not reachable by a path from any mutant vertex, the process never ends in absorption. 
We will thus consider only these starting states:
\begin{defn}
\emph{A starting state} is a state $(G,M)$ such that for every resident vertex $u\in V(G)\setminus M$ there exists a path from some mutant vertex $v$ to the vertex~$u$ (directed path if $G$ is a directed graph).
\end{defn}

If the process begins in a starting state, with probability $1$ it ends in fixation in finite time. 
Note that for undirected graphs, all states are starting states if and only if the graph $G$ is connected (for disconnected graphs, no states with exactly one mutant are starting states). We will often consider starting states that consist of only one mutant vertex.

Similarly to the classical Moran birth-death process with finite $r>1$, we can consider the absorption and fixation times. Because in this particular case, the process ends in fixation with probability one, the absorption and fixation time coincide. Both of them count the expected number of steps until all vertices become mutants and we call this value to be the \emph{colonization time}. We will denote by $\T(G,v)$ the colonization time under the continuous process with starting state $(G,\{v\})$. Similarly, we denote an analogous quantity by $\TL(G,v)$ for the classic Moran process. We will often study the worst-case scenarios in order to prove upper bounds on the quantities $\T(G,v)$ and $\TL(G,v)$. Thus we also denote $\T(G) = \max_{v\in V(G)}\T(G,v)$ the maximum expected colonization time over all starting vertices and similarly $\TL(G) = \max_{v\in V(G)}\TL(G,v)$.

Note that in each step of both of these processes, the new state either remains the same or gains a new mutant. Hence, we are gaining the mutants one by one, and thus we can divide the steps of the process into stages.

\begin{defn}
Let us fix a starting state $(G,\{v\})$. We break the time needed for all the vertices to become mutants into $n-1$ stages. One stage consists of gaining one more mutant, meaning in stage $k$, we start with $k$ mutants and end with $k+1$ mutants. With a slight abuse of notation, we denote the probability of gaining a new mutant in stage $k$ as $P_k$ and the expected time until this happens as $\E[t_k]=\frac{1}{P_k}.$ Formally, this depends not only on $k$ but also on the current state.
\end{defn}

To distinguish the steps in which the state does change from those when it doesn't, we define the notion of an active edge and an active vertex:

\begin{defn}
An \emph{active edge} is an edge between a mutant vertex and a resident vertex.
An active vertex is a vertex that is incident to at least one active edge.
\end{defn}

With these definitions, we can show our first result. It states that the classic Moran process is always the same or faster than the continuous one.

\begin{thm}\label{thm:normal is slower than limit}
For any graph $G$ (directed or undirected) and any vertex $v$ we have $\T(G,v)\geq \TL(G,v).$
\end{thm}
\begin{myproof}
Consider any configuration $X$ of nodes currently occupied by mutants. Denote the number of mutants by $|X|=k$.
Recall that $P_k$ is the probability of gaining a mutant in a single step in the continuous process.
Similarly, for the classic Moran process, denote the probability of gaining a mutant in a single step by $P_k^L$.
We claim that $P_k = \frac{k}{n}\cdot P_k^L$. Indeed, in the continuous process, the probability of picking a particular mutant for reproduction is $\frac{1}{n}$, whereas it is $\frac{1}{k}$ in the classic case.
Plugging in $k\le n$ we find $P_k \leq P_k^L$ and so $\E[t_k^L]\geq \E[t_k]$.
As this holds for any $k$ and any mutant configuration $X$ with $k$ mutants, we get the desired result $\T(G)\geq \TL(G)$.
\end{myproof}

As mentioned earlier we believe the continuous process is more natural. One of the motivations is that this corresponds to the idea of population where the reproduction times of individuals are exponentially distributed random variables. In order to show this connection we first define a notion of a \emph{real time}.

\begin{defn}[Real time]
Consider a graph $G$ and the classical Moran process with mutant fitness 1 and resident fitness 0. We define time units such that one step of the Moran process from state $(G,M)$ accounts for $\frac{1}{|M|}$ units of time (note that in this case, $|M|$ is the total fitness of the population).
We define \emph{real time} to be the expected number of time units until all vertices become mutants. We will denote $\R(G,v)$ to be the real time until fixation starting in state $(G,\{v\}).$
\end{defn}

Next, we can prove that the colonization time is always exactly $n$-times larger than the real time.

\begin{thm}
For any graph $G$ (directed or undirected) and any vertex $v$ we have $\T(G,v)=n\cdot\R(G,v).$ 
\end{thm}
\begin{myproof}
It is sufficient to prove that, in expectation, the time needed for one step of the continuous process is $n$ times larger than the real time. For that let us fix a state $s = (G,M)$ and let us denote the active edges in this state to be $E_s.$ To make one step of the process we need to select an active edge. In the continuous process, this is done with probability $\sum_{(u,v)\in E_s}\frac{1}{n}\cdot\frac{1}{\deg{u}}$ so the expected time of one step is $1/\sum_{(u,v)\in E_s}\frac{1}{n}\cdot\frac{1}{\deg{u}}.$ For the classical Moran process the probability of gaining a mutant is $\sum_{(u,v)\in E_s}\frac{1}{|M|}\cdot\frac{1}{\deg{u}}$, the expected time until it happens is thus $1/\sum_{(u,v)\in E_s}\frac{1}{|M|}\cdot\frac{1}{\deg{u}}.$ By the definition this one step accounts for $\frac{1}{|M|}$ units of the real time. So the expected number of real time units needed to obtain one more mutant is:
$$\frac{1}{|M|}\cdot \frac{1}{\sum_{(u,v)\in E_s}\frac{1}{|M|}\cdot\frac{1}{\deg{u}}} = \frac{1}{\sum_{(u,v)\in E_s}\frac{1}{n}\cdot\frac{1}{\deg{u}}},$$
which is exactly the expected time it takes to get one more mutant in the continuous process and hence the proof is done.
\end{myproof}

\begin{rem}
Throughout this text, we will also use the asymptotic notation $o(\cdot)$, $O(\cdot)$, $\Omega(\cdot)$ and $\Theta(\cdot)$ to denote that some function $f$ is asymptotically strictly smaller than some other function $g$ (denoted $f=o(g)$), asymptotically smaller than or equal to ($f=O(g)$), asymptotically larger than ($f=\Omega(g)$) and asymptotically equal to ($f=\Theta(g)$). We will also use the symbol $\approx$ to denote "approximately equal to," meaning $f(n)\approx g(n)$ if $f(n)=g(n)+o(g(n))$. For example $\frac{1}{2}n^2+3n = o(n^3)=O(n^2)=\Omega(n\log n)=\Theta(n^2).$ For more details, see \cite{arora2009computational}.

We also use $\H_n$ in the later chapters to denote the harmonic number. That is the partial sum of harmonic series, meaning $\H_n = \sum_{k=1}^n\frac{1}{k}\approx\log{n}.$
\end{rem}

\section{General bounds}\label{sec:general}
Here, we prove \cref{thm:upper bound}
 and \cref{thm:backward} from the main text.

\subsection{Upper bounds}
By \cref{thm:normal is slower than limit}, the classic Moran process is always faster (or as fast as) the continuous process. Therefore, in this section, we prove the upper bounds for the continuous process, which proves the bounds for both processes.

\directedupperbound*

\begin{myproof}
We have at least one active vertex in the $k$-th stage and probability $\frac{1}{n}$ to pick this vertex.
Let us denote by $a$ the number of active edges, and by $b$ the number of non-active edges incident to this vertex.
We can have at most $k-1$ non-active edges from this vertex, and so $b\leq k-1$.
The probability of picking an active edge incident to this vertex is thus $\frac{a}{a+b}\geq \frac{a}{a+k-1}\geq\frac{1}{k}$.
In total, that gives us probability $\geq\frac{1}{n\cdot k}$ to gain a mutant in one step.
The expected time until this happens is thus at most $nk$ in the $k$-th stage.
When we sum this over all stages, we get that the expected time is at most $\sum_{k=1}^{n-1}nk=n\cdot\frac{ (n-1)\cdot n}{2}=\frac12 n^3-\frac{1}{2}n^2$.
\end{myproof}


We proved that for every graph $G_n$ on $n$ vertices, the process will take at most $O(n^3)$ time. 
This contrasts with the classical Moran process with finite $r>1$ which can be on some spatial structures exponential. Even more, the bound in~\cref{thm:upper bound} is exactly tight. We identify the slowest population structure that achieves this bound, concretely, the backward graph.

\begin{defn}[Backward graph]
For every $n$, consider a directed graph $\B_n$ defined as follows: vertices are denoted $\{1,2,\dots,n\}$, and there are forward and backward edges. Forward edges are of the form $(i, i+1)$ for every $i\leq n-1$. Backward edges lead between every pair of vertices $(j, i)$ such that $i<j$ (see \cref{fig:upper_bound_digraf}).
\begin{figure}[!hbt]
    \centering
    \includegraphics[scale=0.6]{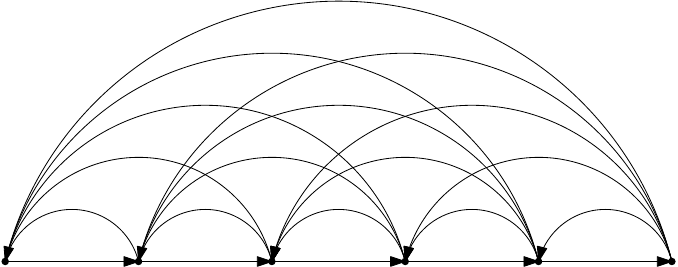}
    \caption{Backward graph $\B_6$.}
    \label{fig:upper_bound_digraf}
\end{figure}
\end{defn}

\backwardgraph*

\begin{myproof}
We set $G_n=\B_n$ and $v$ as the node labeled 1 (the leftmost node).
It is easy to see that in that case, we have to gain the vertices in the order $2,3,\dots, n$.
In the $k$-th stage, we always have one active vertex $k$ with $k-1$ non-active (backward) edges and only one active (forward) edge.
Therefore, in $k$-th stage, the probability of gaining a new mutant is exactly $\frac{1}{nk}$ (that means the bounds used in \cref{thm:upper bound} are satisfied with equality).
And so the expected time until all vertices become mutants is in total exactly $\frac{n^2\cdot (n-1)}{2}=\frac{1}{2}n^3+o(n^2)$.
\end{myproof}

Moreover, we can show that asymptotically, the bound is tight also for DAGs. We use the half-path graph as an example of a DAG with the expected colonization time $O(n^3)$.

\begin{defn}[Half-path graph]
For every even $n=2k$, we define a \emph{half-path} graph (denoted as $\HP_n$) as a DAG with vertices $\{1,2,\dots, n=2k\}$, and edges leading as follows: The edges form a path in the first half of the vertices, meaning $(i,i+1)$ is an edge for $i\in\{1,2,\dots, k-1\}$. The second half is connected to every vertex in the first half, meaning we have an edge $(i,j)$ for all $i\in\{1,2,\dots, k\}$, $j\in\{k+1,k+2,\dots, 2k\}$ (see \cref{fig:upper_bound_dag}).

\begin{figure}[!hbt]
    \centering
    \includegraphics[scale=0.6]{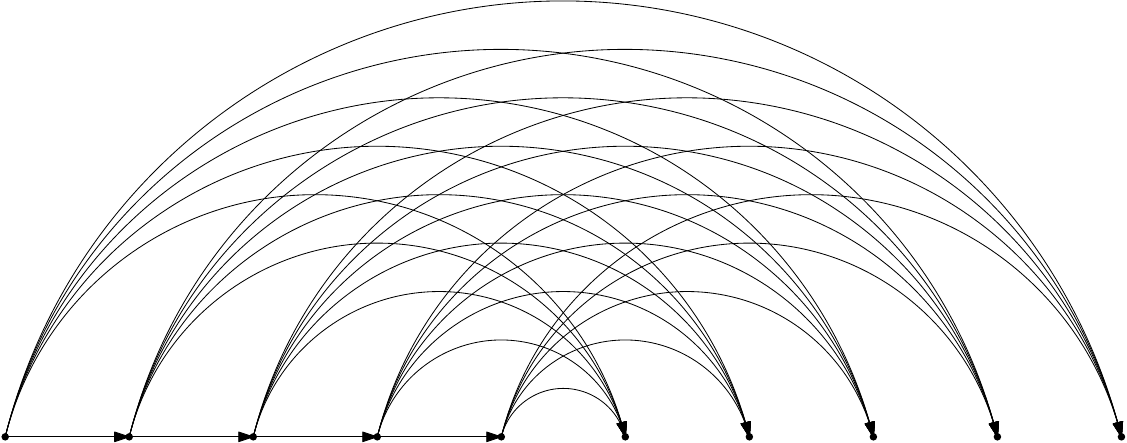}
    \caption{Half-path graph $\HP_{10}$.}
    \label{fig:upper_bound_dag}
\end{figure}
\end{defn}

\begin{thm}
For every even $n$ there exists a directed acyclic graph $G_n$ and an initial mutant node $v$ of $G_n$ such that $\T(G_n,v) =\frac{1}{4}n^3+o(n^2).$  
\end{thm}
\begin{myproof}
We set $G_n=\HP_n$ and $v$ as the node labeled 1 (the leftmost node).
To get to the state where all vertices are mutants, we must follow the path in the first half of the vertices.
This path has length $k-1$, and the probability of gaining one vertex on this path is $\frac{1}{n(k+1)}$, so the expected time is $n(k+1)$.
Altogether the total expected time is at least $n(k+1)(k-1)=n(\frac{n}{2}+1)(\frac{n}{2}-1)=\frac{1}{4}n^3+o(n^2)$.
\end{myproof}

We state two remarks.

First, in the proof, we don't care how long it takes for mutants to claim the second half of the vertices.
Just looking at the first half of the vertices, it already takes $\Omega(n^3)$ time in expectation.

Second, for simplicity, the definition of the half-path graph considers only even~$n$, but it is easy to generalize this also for odd $n$.
For $n=2k+1$, we can do the same construction as for $2k$ and then add the vertex $n+1$ to the second half of the vertices, meaning that we also add the edges of the form $(i,n+1)$ such that $i\leq k$. Following the same arguments proves the same for this modified graph with odd $n$.

\subsection{Lower bound} 

We prove a general lower bound for both directed and undirected graphs.

\begin{thm}
Let $G_n$ be a graph (directed or undirected) with $n$ nodes 
and $v$ any initial mutant node. Then $\T(G_n, v)\geq n\cdot H_{n-1}=\Omega(n\log{n}).$ 
\end{thm}
\begin{myproof}
To gain a new mutant, we must first pick a mutant vertex.
In stage $k$, the probability of picking a mutant is $\frac{k}{n}$ and therefore $P_k\leq \frac{k}{n}$ and so $\E[t_k]\geq \frac{n}{k}$.
When we sum this up over the $n-1$ stages, we get 
$$ \T(G_n)\ge \sum_{k=1}^{n-1}\E[t_k]\geq\sum_{k=1}^{n-1}\frac{n}{k}=n\log n+o(n\log n).$$
\end{myproof}

\section{A stronger bound for regular graphs}\label{sec:regular}
As we showed in the previous section, the expected time can be as large as~$\Omega(n^3)$.
In this section, we prove stronger upper bounds for particular classes of graphs.
First, we look at $d$-regular undirected graphs.

\regular*
\begin{myproof}
The idea of the proof is as follows. We look at the stages in which the mutants conquer the graph. A stage is fast if there are many active edges compared to the regularity constant $d$. Hence, we care only about the slow stages with a few active edges. But at the end of a slow stage, we gain a new mutant such that necessarily most of this mutant's neighbors are residents. As these new active edges are all incident to one mutant vertex, their number can decrease only by one per stage. That means that the number of active edges will be large in the next $\frac{d}{2}$ stages. Hence, we can aggregate the time spent on the one slow stage with the upcoming $\frac{d}{2}$ fast stages and conclude that, on average, we spent linear time per stage, thus giving us the desired bound $O(n^2).$

More precisely, consider stage $k$ and let $e_k$ denote the number of active edges.
As the graph is $d$-regular, the probability of gaining a mutant in stage $k$ is $\frac{e_k}{d\cdot n}$.
The idea is to distinguish two cases:
Either $e_k$ is large, and so the probability of gaining a mutant is large as well;
Or $e_k$ is small, but then the new mutant will have many active edges (see \cref{fig:regular}), so if we look at the expected time needed to gain several new mutants, the process will again be reasonably fast.

\begin{figure}[!hbt]
    \centering
    \includegraphics[scale=0.6]{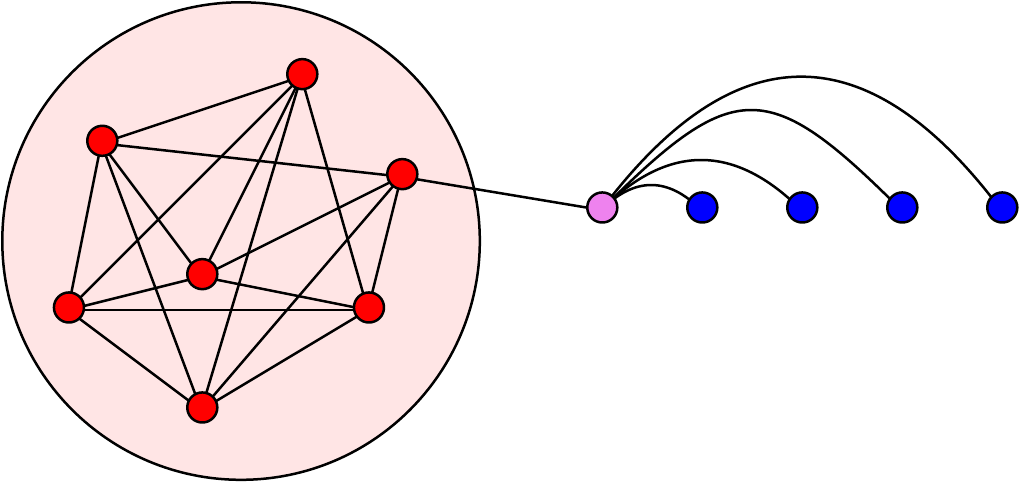}
    \caption{Vizualization of the key idea in the proof. Red vertices represent mutants and blue vertices represent residents. The pink vertex is the soon-to-be infected individual. As a few edges join this node to the mutant part of the graph, there are many outgoing edges to other residents which makes the subsequent phases fast.}
    \label{fig:regular}
\end{figure}

Formally, we distinguish two cases: either $e_k\geq \frac{d}{4}$ or $e_k < \frac{d}{4}$.
\begin{enumerate}
    \item If $e_k\geq \frac{d}{4}$, then the probability of gaining a new mutant is $\frac{e_k}{d\cdot n}\geq \frac{\frac{d}{4}}{d\cdot n}=\frac{1}{4n}$.

    \item If $e_k < \frac{d}{4}$, then $e_{k+1}\geq \frac{3}{4} d$ (because graph $G$ is $d$-regular) and similarly $e_{k+2}\geq \frac{3}{4} d -1$ and so on until $e_{k+\frac{d}{2}}\geq \frac{d}{4}+1$.
    For us it will be sufficient to know that $\forall i\in \{1,2,\dots, \frac{d}{2}\}$ we have $e_{k+i}\geq \frac{d}{4}$.
    Then $\forall i \in \{1,2,\dots, \frac{d}{2}\}$ we get: 
    $$P_{k+i} = \frac{e_{k+i}}{d\cdot n}\geq \frac{\frac{d}{4}}{d\cdot n}=\frac{1}{4n},$$
    and so $\E[t_{k+i}]=\frac{d\cdot n}{e_{k+i}}\leq 4n$.
    Also, as $\frac{d}{4}>e_k\geq 1$ we have $P_k = \frac{e_k}{d\cdot n}\geq \frac{1}{d\cdot n}$ and so $\E[t_k]\leq d\cdot n$. Now we can sum the expected times for the next $\frac{d}{2}+1$ stages as:
    $$\E[t_k]+\sum_{i=1}^{\frac{d}{2}}\E[t_{k+i}] \leq dn + \sum_{i=1}^{\frac{d}{2}}4n=dn+2dn=3dn.$$
\end{enumerate}
Now, we will compare the expected time of gaining a new mutant or mutants in both of these cases with the case when $e_k=\frac{d}{8}$ for every stage $k$. 

\begin{enumerate}
    \item In the first case, if $e_k=\frac{d}{8}$, then $P_k=\frac{\frac{d}{8}}{d\cdot n}=\frac{1}{8n}$ and so $\E[t_k]=8n\geq 4n$ so we see that it is greater or equal to the result in the first case.

    \item In the second case we have $P_{k+i} = \frac{\frac{d}{8}}{dn}=\frac{1}{8n}$ and so $\E[t_{k+i}]=8n$. If we sum this through the $\frac{d}{2}+1$ stages we get $$\sum_{i=0}^{\frac{d}{2}}8n = 4dn+8n \geq 3dn.$$
    So again, we see that the expected time for gaining the next $\frac{d}{2}+1$ mutants would be greater if it were the case $e_k=\frac{d}{8}$ for all $k$.
\end{enumerate}
Altogether, we see that our $d$-regular graph is not slower than if it was the case that $e_k=\frac{d}{8}$ for every stage $k$. But if all $e_k=\frac{d}{8}$, then the stage $k$ has expected time $8n$ and so expected time summed through all stages would be $\sum_{k=1}^{n-1}8n=8n(n-1)=O(n^2).$ And as our $d$-regular graph is not slower than this it follows that $\T(G_n)=O(n^2)$.
\end{myproof}


Note that the constant in the big $O$ doesn't depend on the regularity constant~$d$. Also, in general, this asymptotic bound is tight because as we will show in \cref{thm:cycle} it holds $T(C_n)=O(n^2),$ where $C_n$ is the cycle on $n$ vertices.

\subsection*{Relaxing the regularity condition}
Next, we relax the condition of regularity to allow vertices to have degrees in some fixed range $[d, D]$. 

\begin{thm}
For undirected graph $G$ with minimum degree $d$ and maximum degree~$D$ such that $D\leq c\cdot d$ for some constant $c\geq 1$ it holds $T(G_n)=O(n^2).$
\end{thm}
\begin{myproof}
Very similar to the previous proof. Again, we distinguish two cases $e_k\geq \frac{d}{4}$ and $e_k < \frac{d}{4}$.

\begin{enumerate}
    \item If $e_k\geq \frac{d}{4}$, then the probability of gaining a new mutant is at least \linebreak$\frac{e_k}{D\cdot n}\geq \frac{\frac{d}{4}}{cd\cdot n}=\frac{1}{4cn}$.

    \item In the case $e_k < \frac{d}{4}$ let us denote $d+x$ to be the degree of the mutant gained after $k$--th stage. Clearly $x\geq 0$. Then $e_{k+1}\geq d+x-\frac{d}{4}=\frac{3}{4}d+x.$ and $e_{k+\frac{d}{2}}\geq d+x-\frac{d}{4}-\frac{d}{2}=\frac{d}{4}+x$ and so again we get $\forall i \in \{1,2,\dots, \frac{d}{2}\}$:
    $$P_{k+i} = \frac{e_{k+i}}{(d+x)\cdot n}\geq \frac{\frac{d}{4}+x}{(d+x)\cdot n}\geq\frac{1}{4n}.$$
    In the $k$--th stage, we have at least one active edge incident to the mutant with a degree at most $D$. So $P_k\geq \frac{1}{D\cdot n}$ and $\E[t_k]\leq D\cdot n$. If we again sum this together with $\E[t_{k+i}]\leq 4n$ we get similar sum as before:
     $$\E[t_k]+\sum_{i=1}^{\frac{d}{2}}\E[t_{k+i}] \leq Dn + \sum_{i=1}^{\frac{d}{2}}4n=Dn+2dn\leq cdn+2dn = (c+2)dn.$$
     The average time spent per one stage in this case is thus at most $\frac{(c+2)dn}{\frac{d}{2}+1}.$

\end{enumerate}
Again we can compare it with the situation when $\forall k\in \{1,\dots, n-1\}: e_k = \frac{d}{8c}$. In this case, the probability of gaining a new mutant in one stage is at most $\frac{\frac{d}{8c}}{dn}=\frac{1}{8cn}$. Hence, the expected time until this happens is at least $8cn.$ We want to show that this time is slower than the average time at one stage in the original process. In the first case, we get the inequality $8cn\geq 4cn$, which is always satisfied. In the second case, we want to prove:
$$8cn\geq \frac{(c+2)dn}{\frac{d}{2}+1}.$$
 That is equivalent to $8c\cdot dn+16cn\geq (2c+4)dn$. And as $c\geq 1$, we have $8c\geq 2c+4$, and thus the inequality is satisfied. So again, we see that our graph is not slower than a graph with $e_k=\frac{d}{8c}$ for all $k$. In this setting, the probability of gaining a mutant at each step is at least $\frac{\frac{d}{8c}}{Dn}\geq \frac{1}{8c^2n}$ and the expected time for one stage is thus $8c^2n$. Because the original process is faster than this, we get $\T(G_n)=O(n^2)$.
\end{myproof}

\section{A stronger bound for undirected graphs}\label{sec:undirected}

We have seen that for regular undirected graphs, we can get an upper bound of $O(n^2).$ However, this cannot be generalized for all undirected graphs because, as we prove in \cref{thm:star}, there exists an undirected graph, notably the star, on which the continuous process takes $\Omega(n^2\log n)$ time in expectation. We prove a slightly weaker bound here.  

\undirected*

\begin{myproof}
Let $d=\sqrt n$ be an auxiliary threshold value. Let us call any vertex with degree $\geq d$ a \emph{large vertex}. Vertices that are not large are called \emph{small}.
We will prove that, on average, one stage takes roughly $n\sqrt{n}$ steps (up to a constant).

Formally, consider stage $k$ and let $e_k$ denote the number of active edges. We distinguish several cases.
\begin{enumerate}
    \item Suppose that $e_k\geq \frac{1}{4}d$. For each active edge, the probability that the reproduction event happens along precisely that edge is at least $\frac1n\cdot\frac1{n-1}\ge \frac1{n^2}$.
    Thus, the probability of gaining a new mutant in the next step is at least $P_k\ge\frac{\frac{1}{4}d}{n^2}=\frac{d}{4n^2}$ and the expected time until this happens is $\E[t_k]\le \frac{4n^2}{d}= 4n\sqrt n$, where in the inequality we used $d\ge \sqrt n$.
    \item Suppose that $e_k<\frac{1}{4} d$. Then, we have three sub-cases depending on the situation of large resident vertices.
    \begin{enumerate}
        \item \emph{There exists a large resident vertex with a mutant neighbor.}
        
        Call the large resident vertex $v$ and its mutant neighbor $u$.
        Since $e_k<\frac14d$, at least $d-e_k>\frac34d$ of $v$'s neighbors are residents.
        We will prove that the expected time per stage over the next $1+\frac12d$ stages is at most $2n\sqrt n$.
        
        First, we wait until $v$ becomes mutant. The probability that edge $(u,v)$ is selected for reproduction is at least $1/n^2$, which takes at most $n^2$ steps in expectation.
        
        Second, once $v$ becomes a mutant, we wait until at least $\frac34d$ of $v$'s neighbors are mutants. Note that some of them might have become mutants while we were waiting for $v$.
        Also, note that once $\frac34d$ of $v$'s neighbors are mutants, we have indeed gained at least $1+\frac34d-\frac14d=1+\frac12d$ mutants.
        As long as at most $\frac34d$ of $v$'s neighbors are mutants, at least $d-\frac34d=\frac14d$ of $v$'s neighbors are residents, and therefore there are at least $\frac{1}{4}d$ active edges incident to $v$.
        The probability of gaining a mutant in one step through one of these edges is thus at least $\frac1n\cdot\frac{\frac14d}{d}=\frac{1}{4n}$, and so the expected time is at most $4n$.
        In total, over the next $1+\frac{1}{2}d$ stages we spend at most $1\cdot n^2+(\frac12d)\cdot 4n=n^2+2nd$ steps. The average time per stage is thus at most $$\frac{n^2+2nd}{\frac{1}{2}d+1}\leq \frac{nd^2+2nd}{\frac{1}{2}d+1}=\frac{(\frac{1}{2}d+1)(2nd)}{\frac{1}{2}d+1}=2nd\le 2n\sqrt n,$$

        where in the two inequalities, we used $n\le d^2$ and $d\le \sqrt n$.
        
        \item \emph{Case {\normalfont (a)} does not occur, and there exists a large resident vertex somewhere in the graph.}

        We find and fix some shortest path $v_0, v_1, v_2, \dots, v_l$ between a mutant vertex and some large resident vertex. 
        That is, $v_0$ is a mutant vertex, $v_1,v_2,\dots, v_{l-1}$ are small resident vertices, and $v_{l}$ is a large resident vertex.
        Then we wait until $v_1,\dots,v_l$ and at least half of $v_l$ neighbors become mutants.
        Note that since we consider the shortest path, vertex~$v_l$ initially has no mutant neighbors. 
        
        We proceed as in case (a).
        First, node $v_1$ becomes mutant in at most~$n^2$ steps in expectation. 
        Since vertices $v_1,\dots,v_{l-1}$ are small, each vertex $v_2,\dots,v_l$ becomes mutant in at most $1/(\frac1n\cdot\frac1d)=nd$ steps in expectation.
        Once the large vertex $v_l$ becomes mutant, we wait until at least $d/2$ of its neighbors are mutants. Using the same method as in (a), we aggregate the first ``slow'' stage (in which $v_1$ becomes mutant) and the remaining $(l-1)+\frac12d\ge \frac12d$ ``fast'' stages. By the same algebra as in~(a), the average time per stage will again be at most $2nd\le 2n\sqrt n$ in expectation.
        
        \item \emph{Cases {\normalfont (a)} and {\normalfont (b)} do not occur; that is, all resident vertices are small.}
        
        Once this happens, we aggregate the time spent from this moment on until all vertices become mutants. 
        Therefore, this case happens only once.
        Suppose it happens at stage $k$.
        We know that at this point $e_k<\frac{1}{4}d$.
        First, we wait until all these $e_k$ edges are used for reproduction.
        Each one of them is used after at most $n^2$ steps in expectations, so in total, we wait at most $e_k\cdot n^2\leq\frac14n^2d\le \frac14n^2\sqrt n$ steps in expectation.
        From that point on, all active vertices will be small vertices.
        As we computed in (b), the expected time until we use an edge from a small vertex is at most $nd\le n\sqrt n$.
    \end{enumerate}
\end{enumerate}
In each case, we spend, on average, at most $\max(4n\sqrt n,2n\sqrt n,n\sqrt n)$ steps per one stage.
Plus, once during the process, we wait at most $\frac14n^2\sqrt n$ steps in case (c).
Since there are $n-1<n$ stages, the total expected time is at most $(n-1)\cdot 4n\sqrt{n}+\frac{1}{4}n^2\sqrt{n}=4n^2\sqrt{n}+o(n^2\sqrt{n})=O(n^2\sqrt{n}).$
\end{myproof}

\section{Specific graphs}\label{sec:specific}

In this chapter, we show how fast is the continuous process for some specific graphs. We analyze cycles, cliques, star graphs, double stars, and total order graphs.

\subsection{Cycle}

\begin{thm}[Cycle]\label{thm:cycle}
Let $\C_n$ be a cycle with $n$ nodes. Then $\T(\C_n)=\Theta(n^2).$
\end{thm}
\begin{myproof}
At each stage, the mutants cover a consecutive segment of the cycle. Hence, at every time, exactly two active edges connect the ends of the mutant segment with the rest of the cycle. To pick an active edge, we need to select the incident mutant, which is done with probability $\frac{1}{n}$, and then pick this particular edge. But because the degree of each node is $2$, this has probability $\frac{1}{2}.$ The two active edges are thus picked with probability $2\cdot\frac{1}{n}\cdot\frac{1}{2}=\frac{1}{n}$ independently of the stage number. The expected time until we gain a new mutant is then always~$n$. Because we have $n-1$ stages, in total the expected time is $n(n-1)=\Theta(n^2).$ 
\end{myproof}

\subsection{Clique}

\begin{thm}[Complete graph]\label{thm:complete}
Let $\K_n$ be a complete graph with $n$ nodes.
Then $\T(\K_n)=2(n-1)\H_{n-1} = \Theta(n\log n)$.
\end{thm}
\begin{myproof}
In the $k$-th stage, we have the probability $\frac{k}{n}$ of picking a mutant vertex. Every mutant vertex then has $n-k$ active edges. Hence, the probability of picking one is $\frac{n-k}{n-1}$. In the $k$-th stage, the probability of gaining a mutant is thus $\frac{k\cdot (n-k)}{n\cdot(n-1)}$ and the expected time until this happens is $\E[t_k] = \frac{n\cdot(n-1)}{k\cdot (n-k)}=(n-1)\cdot(\frac{1}{k}+\frac{1}{n-k})$. 
Summing this over all stages, we get:
$$\T(\K_n)=\sum_{k=1}^{n-1} (n-1)\cdot\left(\frac{1}{k}+\frac{1}{n-k}\right) = (n-1)\cdot2\cdot\sum_{k=1}^{n-1}\frac{1}{k}=2n\log n+o(n\log n).$$
\end{myproof}

\subsection{Star and double star graphs}

\begin{figure}[!hbt]
    \centering
    \begin{subfigure}{.4\textwidth}
      \centering
      \includegraphics[scale=0.7]{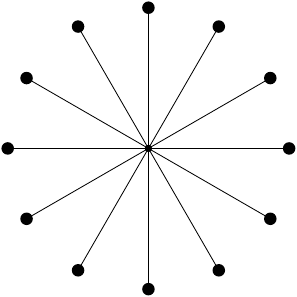}
      \caption{Star graph $\S_n$}
      \label{fig:star}
    \end{subfigure}
    \hspace{.01\textwidth}
    \begin{subfigure}{.4\textwidth}
      \centering
      \includegraphics[scale=0.6]{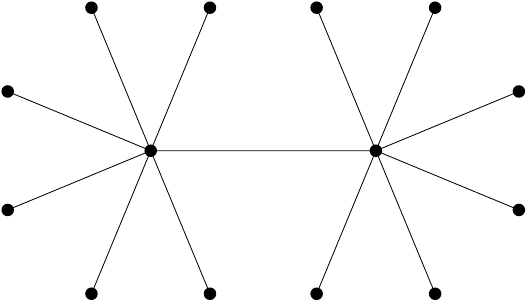}
      \caption{Double star graph $\D_n$}
      \label{fig:doble_star}
    \end{subfigure}\\
    \vspace{1.0cm}
    \begin{subfigure}{.4\textwidth}
      \centering
      \includegraphics[scale=0.7]{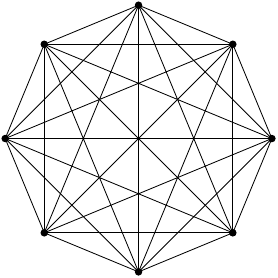}
      \caption{Complete graph $\K_n$}
    \end{subfigure}%
    \hspace{.01\textwidth}
    \begin{subfigure}{.4\textwidth}
      \centering
      \includegraphics[scale=0.5]{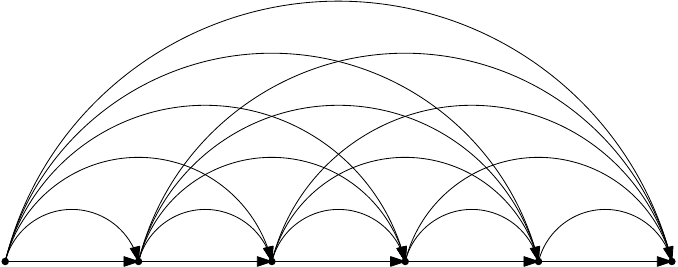}
      \caption{Total order graph $\TO_n$}
      \label{fig:total_order}
    \end{subfigure}\\
    \caption{Specific graphs.}
    \label{fig:graphs_defn}
\end{figure}

\begin{defn}[Star graph]
For every $n=k+1$, we define a \emph{star} graph (denoted as $\S_n$) as an undirected graph with vertices $\{1,2,\dots, n=k+1\}$. We call the vertex $k+1$ to be the center and all the other vertices to be leaves. The edges connect all leaves to the center, meaning that for every $i\in \{1,\dots, k\}$ the pair $(i,k+1)$ is an edge (see \cref{fig:star}).

\end{defn}

The double star graph can be obtained by gluing two copies of the same star graph together.

\begin{defn}[Double star graph]
For every even $n=2k+2\geq 4$, we define a \emph{double star} graph (denoted as $\D_n$) as an undirected graph with vertices $\{1,2,\dots, n=2k+2\}$. This graph is obtained by taking two star graphs, one with center $k+1$ and leaves $1,2,\dots, k$ and the other one with center $2k+2$ and leaves $k+2,\dots, 2k+1$ and joining them by an edge connecting their centers $(k+1,2k+2)$ (see \cref{fig:doble_star}).

\end{defn}

Next, we prove the asymptotic colonization times on star and double star graphs. Note that the star graph is the slowest undirected graph we found for the continuous process. We also prove that the double star is asymptotically as fast as the star. However, this is not true for the Moran process with finite $r>1$. In the finite case, it can be proven that the star takes $\Theta(n^2\log n)$ time and double star at least~$\Omega(n^3).$

\begin{thm}[Star graph]\label{thm:star}
For a star graph $\S_n$ with $n$ nodes and any initial mutant node $v$ we have $\T(\S_n,v)=\Theta(n^2\log n).$
\end{thm}
\begin{myproof}
The first stage takes $n$ steps on average.
After that, it is always the case that mutants occupy the center of the star and one leaf, no matter where the initial mutant started.
In the $k$-th stage ($k\ge 2$), there are $k-1$ mutant leaves and $n-k$ non-mutant leaves.
The probability of gaining a new vertex is therefore $\frac{1}{n}\cdot \frac{n-k}{n-1}$. The expected time until this happens is thus $\frac{n\cdot (n-1)}{n-k}$. If we sum this up over all the stages, we get the following:

\begin{equation*}
\begin{split}
\T(\S_n) &=  \sum_{k=1}^{n-1} \frac{n\cdot (n-1)}{n-k} =n\cdot(n-1)\cdot \sum_{k=1}^{n-1}\frac{1}{n-k}\\
&=n\cdot(n-1)\cdot \sum_{k=1}^{n-1}\frac{1}{k}=n^2\log n + o(n^2\log n).
\end{split}
\end{equation*}
\end{myproof}


\begin{thm}[Double star]\label{thm:doublestar}
For a double star graph $\D_n$ with $n=2k+2\geq 4$ nodes and any initial mutant node $v$ we have $\T(\D_n,v)=\Theta(n^2\log n).$
\end{thm}
\begin{myproof}
Let us denote the two centers of the double star as $x_1$ and $x_2$. WLOG, we start in the left half of the graph. 
Let us divide the process into two phases. 

The first phase will end when we gain the vertex $x_2$, and the rest of the process will be the second phase. The expected time of the first phase is surely $\leq \frac{n}{k+2}\cdot\T(\S_{k+2})\leq 2\cdot\T(\S_{n})=\Theta(n^2\log n),$ because we can look at the left star with center~$x_1$ and its neighbors (including $x_2$) as a star graph with $k+2$ vertices. The constant~$\frac{n}{k+2}$ is there to adjust to the fact that the probability of picking a particular vertex in the double star is $\frac{1}{n}$ whilst it is $\frac{1}{k+2}$ in the star $\S_{k+2}.$

\begin{figure}[!hbt]
    \centering
    \includegraphics[scale=0.7]{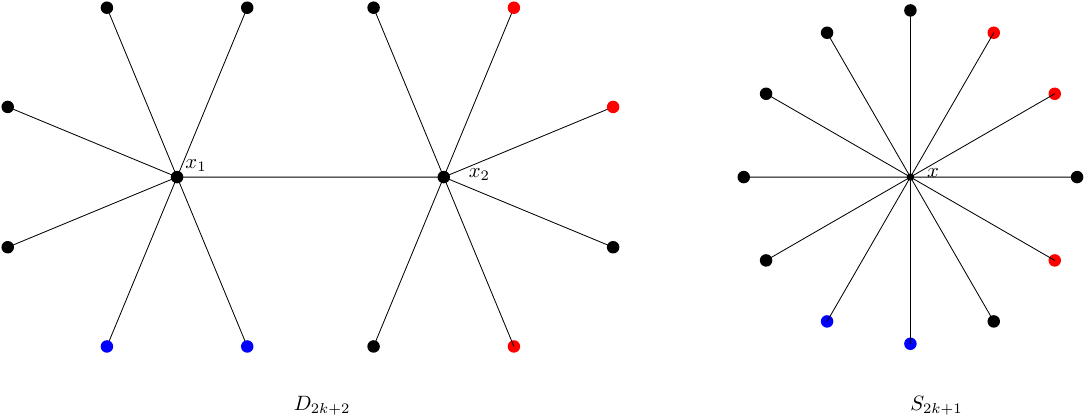}
    \caption{A double star $\D_{2k+2}$ and its comparison to star $\S_{2k+1}$ in the second phase.}
    \label{fig:double_star}
\end{figure}

In every stage in the second phase, the vertices $x_1$ and $x_2$ are already mutants and also some leaves. Suppose there are $a$ mutant leaves next to $x_1$ and $b$ mutant leaves next to $x_2$. We want to compare this process with the process on a star graph $\S_{2k+1}$ with mutant center $x$ and $a+b$ mutant leaves. Then, the probability of gaining one more mutant in the next step in the double star case is: 
$$\frac{1}{n}\cdot\frac{k-a}{k+1}+\frac{1}{n}\cdot\frac{k-b}{k+1}=\frac{2k-a-b}{n(k+1)}.$$
On the other hand in the star graph $\S_{2k+1}$ in stage $m+n+1$ the probability is:
$$\frac{1}{n-1}\cdot\frac{2k-a-b}{2k}.$$
We would like to know when the first probability is larger, for that we can compare only the denominators:
\begin{align*} 
            \frac{1}{n(k+1)}&\geq \frac{1}{(n-1)\cdot 2k}\\
            (n-1)\cdot 2k&\geq n(k+1)\\
            2nk-2k&\geq nk+n\\
            nk-2k-n &\geq 0\\
            (n-2)\cdot(k-1)=2k\cdot(k-1)&\geq 2
\end{align*}
The last inequality is true whenever $k>1$. The probability of gaining a mutant in the double--star case is thus always greater than the probability of gaining a mutant in the star graph. So altogether, we see that the second phase has expected time at most $\T(\S_{2k+1})=\T(\S_{n-1})=\Theta(n^2\log n)$.
Both phases of the double star graph together thus have expected time $O(n^2\log n).$

For the lower bound, it is sufficient to notice that the double star contains the star graph $\S_{k+2}$ as a subgraph (with center $x_1$ and its neighbors) as we did in the first phase above. The colonization time of the double star is thus at least the expected time until mutants gain this subgraph $\S_{k+2}$. This we already computed in the upper bound of the first phase, and it is $\frac{n}{k+2}\cdot \T(\S_{k+2})=\Theta(n^2\log n).$
\end{myproof}

\subsection{Total order graph}

\begin{defn}[Total order graph]
For every $n$, we define a \emph{total order} graph (denoted as $\TO_n$) as a directed graph with vertices $\{1,2,\dots, n\}$ and edges $(i,j)$ for every $i, j\in \{1,2,\dots, n\}$ such that $i<j$ (see \cref{fig:total_order}).

\end{defn}

\begin{thm}[Total order]\label{thm:total-order}
For a total order graph $\TO_n$ with $n$ nodes and a mutant node $v$ being the first node we have $\T(\TO_n,v)=\Theta(n^2).$
\end{thm}

\begin{myproof}
First, we prove the lower bound $\T(\TO_n)=\Omega(n^2)$.
The probability that in a single step, the first resident vertex becomes mutant is $\frac{1}{n\cdot(n-1)}$. The expected time until this happens is $n\cdot(n-1)=\Omega(n^2)$. Thus $\T(\TO_n)=\Omega(n^2)$ as required.

Second, we prove the upper bound $\T(\TO_n)=O(n^2)$.
It is sufficient to prove the bound for $n$ of the form $n=2^k$, so suppose our $n$ is a power of 2.
We divide our $n$ vertices into $k+1$ blocks from left to right, and we index them from zero (as block $0,1,2,\dots, k$).
The sizes of the blocks will be $1,1,2,2^2,2^3,\dots, 2^{k-1}$.
We begin with the first vertex; therefore, the zero block is already mutant. In step~$i$, we wait until the vertices in the $i$-th block become mutants.
There are~$2^{i-1}$ vertices in the block $i$ and $2^{i-1}$ vertices in the previous blocks $0,1\dots, i-1$. Therefore, before any mutant in the $i$-th block is present, there are $2^{i-1}\cdot 2^{i-1}$ active edges incident to $i$-th block.
The probability of using one particular edge is~at least $\frac{1}{n^2}$ and so the probability of gaining one vertex from the $i$-th block is at least~$\frac{2^{i-1}\cdot 2^{i-1}}{n^2}.$ After gaining one mutant we have $2^{i-1}\cdot (2^{i-1}-1)$ active edges incident to $i$-th block and generally after gaining $m$ mutants this number of active edges is $2^{i-1}\cdot(2^{i-1}-m).$ The probability of gaining one more mutant in this situation is thus at least $\frac{2^{i-1}\cdot(2^{i-1}-m)}{n^2}$ so the expected time until this happens is at most $\frac{n^2}{2^{i-1}\cdot(2^{i-1}-m)}.$ Together the expected time until all vertices in the $i$-th block become mutants is then at most:
\begin{align*} 
\sum_{m=0}^{2^{i-1}-1}\frac{n^2}{2^{i-1}\cdot(2^{i-1}-m)}&=\frac{n^2}{2^{i-1}}\cdot\sum_{m=0}^{2^{i-1}-1} \frac{1}{(2^{i-1}-m)}=\frac{n^2}{2^{i-1}}\cdot\sum_{m=1}^{2^{i-1}} \frac{1}{m}= \\
&= \frac{n^2}{2^{i-1}}\cdot \H_{2^{i-1}}\approx\frac{n^2}{2^{i-1}}\cdot(i-1)
\end{align*}
And if we sum this over all the blocks, the expected time until all vertices become mutants is at most:
$$\sum_{i=1}^k \frac{n^2}{2^{i-1}}\cdot(i-1)=n^2\cdot\sum_{i=0}^{k-1} \frac{i}{2^i}\leq n^2\cdot\sum_{i=0}^{\infty} \frac{i}{2^i}=2n^2 = O(n^2).$$
\end{myproof}

\section{Comparison of the two notions of time}\label{sec:comparison}

In this chapter, we show that the two notions of time (times in classic Moran process and in the continuous one) are substantially different.
In particular, we prove that on some graphs, the continuous time is asymptotically larger than the classic Moran one.
Moreover, we show that the classic Moran time is not monotone in the sense described below.
We interpret this as an indication that perhaps the continuous time is more natural.

To illustrate these two claims, we use the lollipop graph as an example.

\begin{defn}[Lollipop graph]
For every $n$, consider an undirected graph $\L_n$ defined as follows: vertices are denoted $\{1,2,\dots,n\}$, and there are two parts of the graph. The first part consists of vertices $1,2,\dots, \sqrt{n}$ and a path between them, meaning there are edges $(i,i+1); \forall i\in \{1,2,\dots,\sqrt{n}-1\}$. The second part of the graph is a clique, meaning there are edges $(i,j); \forall i,j\in\{\sqrt{n}+1,\dots,n\}$. These two parts are then connected by an edge $(\sqrt{n},\sqrt{n}+1
)$ (see \cref{fig:lollipop}).
\begin{figure}[!hbt]
    \centering
    \includegraphics[scale=0.9]{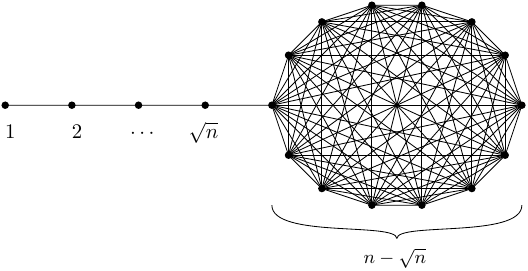}
    \caption{A lollipop graph $\L_n$ on $n$ nodes consists of a path of length $\sqrt n$ that is connected to a clique of size $n-\sqrt{n}$.}
    \label{fig:lollipop}
\end{figure}
\end{defn}

\subsection{The two notions of time are different}

First, we show that the classic Moran time and the continuous time are substantially different on this lollipop graph. The intuition is that when we fix a particular state with $k$ mutants, depending on the value of $k$ the two processes become more similar as $k$ gets larger. That is for the continuous process the probability of picking a mutant in stage $k$ is $\frac{1}{n}$ in comparison with $\frac{1}{k}$ for the classic Moran process. Hence, if $k$ is large (of order $n$), then these processes have probabilities scaled just by some constant, but when $k$ is small (sublinear to $n$) there is an asymptotic difference between these probabilities. Hence, the classic Moran process can be asymptotically faster in these steps. With this intuition, the lollipop graph is chosen exactly in a way to ensure this as we will see in the proof of the following theorem.

\begin{thm}\label{different times}
There exists a graph $G_n$ and an initial mutant node $v$ of $G_n$ such that $\T(G_n, v) = \Theta(n\sqrt{n})$ while $\TL(G_n,v) = \Theta(n\log{n}).$
\end{thm}
\begin{myproof}
We set $G_n=\L_n$ and $v$ as the node labeled 1 (the leftmost node).
To get to the state where all vertices are mutants, we must first follow the path and then spread through the whole clique.

First, consider the continuous process.
Vertex $i$ in the first part will propagate with probability $\frac{1}{2n}$ if $i\neq 1$ and with probability $\frac{1}{n}$ if $i=1$.
Thus, the expected time until the node $\sqrt{n}+1$ becomes mutant is 
$n+\sum_{i=2}^{\sqrt{n}}2n = 2n\sqrt{n}-n.$ Then, similarly to \cref{thm:complete}, we can compute the expected time until all nodes become mutants.
When there are $k$ mutants already in the clique, the probability of gaining a new one is $\frac{k}{n}\cdot\frac{n-\sqrt{n}-k}{n-\sqrt{n}-1}$, thus the expected time until this happens is $\frac{n}{k}\cdot\frac{n-\sqrt{n}-1}{n-\sqrt{n}-k} = n\cdot\frac{n-\sqrt{n}-1}{n-\sqrt{n}}\cdot \left(\frac{1}{k}+\frac{1}{n-\sqrt{n}-k}\right).$
Summing this over all $k$ we get:

\begin{align*}
n\cdot\frac{n-\sqrt{n}-1}{n-\sqrt{n}}\cdot\sum_{k=1}^{n-\sqrt{n}-1} \left(\frac{1}{k}+\frac{1}{n-\sqrt{n}-k}\right) &= \\ 
=n\cdot\frac{n-\sqrt{n}-1}{n-\sqrt{n}}\cdot 2\H_{n-\sqrt{n}-1}&\approx 2n\log{n}.
\end{align*}

For the continuous process, the first part of the graph thus takes $\Theta(n\sqrt{n})$ steps, and the second part takes $\Theta(n\log{n})$ steps in expectation.
Altogether, the process takes $\Theta(n\sqrt{n})$ steps in expectation as claimed.

For the classic Moran process, we proceed similarly.
In the first part, vertex $i$ will propagate with probability $\frac{1}{2i}$ if $i\neq 1$ and with probability $1$ if $i=1$.
Thus, in total the time until the node $\sqrt{n}+1$ becomes mutant is in expectation $1+\sum_{i=2}^{\sqrt{n}}2i = \sqrt{n}\cdot(\sqrt{n}-1)-1 = n-\sqrt{n}-1 = \Theta(n).$ 

In the second part, the expected time can be computed in exactly the same way as for the continuous process, only with the difference that when we have $k$ mutants in the clique already, the probability of picking a mutant node in the clique is $\frac{k}{\sqrt{n}+k}$ instead of $\frac{k}{n}$.
Following the same algebraic modifications, we get that the total expected time for the classic Moran process in the second part is
\begin{align*}
\frac{n-\sqrt{n}-1}{n-\sqrt{n}}\cdot\sum_{k=1}^{n-\sqrt{n}-1} (\sqrt{n}+k)\cdot\left(\frac{1}{k}+\frac{1}{n-\sqrt{n}-k}\right) &= \\
\frac{n-\sqrt{n}-1}{n-\sqrt{n}}\cdot\left(2\sqrt{n}\cdot \H_{n-\sqrt{n}-1} +\sum_{k=1}^{n-\sqrt{n}-1} \left(1+\frac{k}{n-\sqrt{n}-k} \right)\right) &= \\ 
\frac{n-\sqrt{n}-1}{n-\sqrt{n}}\cdot\left(2\sqrt{n}\cdot \H_{n-\sqrt{n}-1} +\sum_{k=1}^{n-\sqrt{n}-1}\frac{n-\sqrt{n}}{k} \right) &= \\ 
\frac{n-\sqrt{n}-1}{n-\sqrt{n}}\cdot (n+\sqrt{n})\cdot \H_{n-\sqrt{n}-1} &= \\ 
&\approx n\log{n}.
\end{align*}
For the classic Moran process, the first part of the graph thus takes $\Theta(n)$ step and the second part takes $\Theta(n\log{n})$ steps in expectation.
Altogether, the classic Moran process takes $\Theta(n\log{n})$ steps in expectation as claimed.
\end{myproof}

\subsection{Classic Moran time doesn't have to be monotone}\label{sec:notmonotone}

Next, we prove that the classic Moran time is not necessarily monotone. That is, we show that if we add additional mutants to the starting configuration, the expected time until fixation might increase, even asymptotically. 
When considering fixation probability rather than the absorption time, this kind of monotonicity is sometimes called \emph{subset domination} and it is known that it is satisfied by the Moran process with finite $r>1$ on undirected graphs \cite{diaz2016absorption}.

\begin{thm} 
There exists a directed graph $G_n,$ an initial mutant node $v$ of $G_n$ and a subset of vertices $\{v\}\subseteq X\subseteq V(G_n)$ such that $\TL(G_n, v) = \Theta(n\log{n})$ while $\TL(G_n,X) = \Theta(n\sqrt{n})$.
\end{thm}

\begin{myproof}
Let us consider a directed lollipop graph $\L'_n$, which is obtained from~$\L_n$ by orienting the edges in the first part from left to right, meaning we will have directed edges $(i,i+1);\forall i\in \{1,2,\dots, \sqrt{n}\}$ on the path. The edges in the clique remain undirected.
Then we set $G_n = \L'_n,$, the starting node $v$ as the node labeled~1 (the leftmost node), and $X = \{\sqrt{n}+1,\dots, n\}\cup \{v\}$.
In other words, $X$ is a set containing all vertices of the clique and the starting node.

It is easy to see that $\TL(\L'_n,v)$ is almost the same as $\TL(\L_n,v)$, which we computed in the proof of \cref{different times}. The only difference is that along the path we gain a new vertex with probability $\frac{1}{n}$ instead of $\frac{1}{2n}$. Asymptotically this is the same. 
Hence $\TL(\L'_n,v)=\Theta(n\log n).$

Now let us look at $\TL(\L'_n, X).$
Since no edge is going out of the mutant clique, the only way to turn
all vertices into mutants is to follow the path in the first half of the graph.
As the out-degree of the nodes in this path is one, to infect the $(i+1)$-th node on this path, we need to pick its mutant predecessor $i$.
If node~$i$ is the last mutant along the path, it is picked with probability $\frac{1}{n-\sqrt{n}+i}$. Thus the expected time until it happens is $n-\sqrt{n}+i.$
Summing this over all vertices on the path, we get that the total time until all vertices become mutants in this setting is:
$$\TL(\L'_n,X) = \sum_{i=2}^{\sqrt{n}} (n-\sqrt{n}+i) = (n-\sqrt{n})\cdot (\sqrt{n}-1)+\frac{\sqrt{n}\cdot(\sqrt{n}+1)}{2}-1 = \Theta(n\sqrt{n}).$$
\end{myproof}

Let us remark that considering the directed lollipop, starting with the clique being mutant using the classic Moran process is asymptotically as fast as considering the undirected lollipop and using the continuous process with only one mutant vertex.
That makes sense because, for the continuous case, the first part of the lollipop was the asymptotically slowest part. And as we previously observed the probabilities of picking a mutant vertex in $k$-th stage between those two processes become more similar as $k$ gets larger.
As we start with $n-\sqrt{n}+1$ mutants in the classic Moran case, these probabilities are already quite similar.



\section*{Acknowledgements}
J.T. and L.K. were supported by the project PRIMUS/24/SCI/012 from Charles University.


\end{document}